\documentclass[pdflatex,sn-aps]{sn-jnl}
\usepackage{graphicx} 
\usepackage{multirow}%
\usepackage{amsmath,amssymb,amsfonts}%
\usepackage{amsthm}%
\usepackage{mathrsfs}%
\usepackage{xcolor}%
\usepackage{textcomp}%
\usepackage{manyfoot}%
\usepackage{booktabs}%
\usepackage{algorithm}%
\usepackage{algorithmicx}%
\usepackage{algpseudocode}%
\usepackage{listings}\usepackage{rotating} 
  \usepackage{tcolorbox}
\usepackage[
singlelinecheck=false 
]{caption}

\usepackage{etoolbox}   
\makeatletter
\patchcmd{\@maketitle}{\artauthors}{\centering{\artauthors}}{}{}
\makeatother

\begin{document}
\title{\vskip -2.8cm Higgs factory options for CERN}
\subtitle{\vskip -0.2cm \large A comparative study}
\date{21 June 2023}

\author[1]{\fnm{A.} \sur{Blondel}}
\author[2]{\fnm{C.} \sur{Grojean}}
\author*[3]{\fnm{P.} \sur{Janot}}\email{patrick.janot@cern.ch}
\author[4]{\fnm{G.} \sur{Wilkinson}}

\affil[1]{\small \orgname{LPNHE Paris-Sorbonne}, \orgaddress{\street{4 place Jussieu}, \city{Paris}, \postcode{75252}, \country{France}}}

\affil[2]{\small Deutsches Elektronen-Synchrotron DESY, Notkestr. 85, 22607 Hamburg, Germany.
}

\affil[3]{\small \orgname{CERN}, \orgdiv{EP Department}, \orgaddress{\street{1 Esplanade des Particules}, \city{Meyrin}, \country{Switzerland}}}

\affil[4]{\small \orgname{University of Oxford}, \orgdiv{Department of Physics}, \orgaddress{\city{Oxford}, \country{United Kingdom}}}

\abstract
{``{\it All future $e^+e^-$ Higgs factories have similar reach for the precise measurement of the Higgs boson properties.}'': this popular statement has often led to the impression that all $\rm e^+e^-$ options are scientifically equivalent when it comes to choosing the future post-LHC collider at CERN. More recently, the concept of sustainability has been added in attempts to rank Higgs factories. 
A comparative analysis of the data currently available is performed in this note to clarify these issues for three different options: the future circular colliders (FCC), and two linear collider alternatives (CLIC and ILC@CERN). 

The main observation is as follows. For the precise measurement of already demonstrated Higgs decays (b\=b, $\tau^+\tau^-$, gg, ZZ, WW) and for $\rm H \to c\bar c$, it would take half a century to CLIC and ILC@CERN to reach the precisions that FCC-ee can achieve in 8 years thanks to its large luminosity and its four interactions points. The corresponding electricity consumption, cost and carbon footprint would also be very significantly larger with linear colliders than with FCC-ee. 

Considering in addition that {\it (i)} FCC-ee is the only place to attempt the measurement of the electron Yukawa coupling, thanks to the ability to produce the Higgs boson directly at $\sqrt{s} = m_{\rm H}$ with reduced centre-of-mass energy spread; {\it (ii)} for the precise measurement of the many Higgs boson couplings that require the production of billions of Higgs bosons (such as H$\gamma\gamma$, HZ$\gamma$, H$\mu\mu$, or HHH), the combination of FCC-ee and FCC-hh is order of magnitude better than what linear colliders can ever do; {\it (iii)} FCC-ee is much more than a Higgs factory and, in an entirely new context where neither the mass scale of new physics nor the intensity of its couplings to the Standard Model are known, only the large luminosities of FCC-ee at the electroweak scale and the parton-parton collision energies at FCC-hh can provide the necessary exploration breadth with a real chance of discovery; and {\it (iv)} the vast experimental programme achievable with both FCC-ee and FCC-hh is out of reach of linear colliders; it is found that FCC-ee is a vastly superior option for CERN, and the only first step en route to the 100\,TeV hadron collider.
}
\maketitle

\tableofcontents

\vfill\eject 

\section{Preamble}

Sustainable development requires an integrated approach that takes into consideration environmental concerns along with economic development.  In 1987, the United Nations Brundtland Commission defined sustainability as ``{\it meeting the needs of the present wiƒthout compromising the ability of future generations to meet their own needs}''~\cite{Brundtland}. 
Similarly, sustainable collider science requires an integrated approach that takes into consideration environmental concerns along with scientific outcome. Indeed, should the scientific outcome be excluded from the concept of sustainability, the best collider would simply be no collider at all, and this would compromise the ability of future generations to make any progress in fundamental physics.  
In our times of measurable climate change due to a global warming of unprecedented rapidity, resulting from the emission of greenhouse gases, a most important criterion in the choice of a future collider at CERN is therefore that its carbon footprint, {\it for a given scientific outcome}, be minimised. Other essential criteria include the corresponding duration of operation, electricity consumption and overall cost of the facility. The acceptance by the the high-energy physics community, the public and the governments will be highly facilitated if it can be shown that all these concerns have taken an important place in the choice.

\section{Introduction}

In January 2020, the European Strategy Group (ESG) stated that {\it an electron-positron Higgs factory is the highest-priority next collider}~\cite{CERN-ESU-015}. In addition, the ESG concluded that {\it Europe should investigate the technical and financial feasibility of a future hadron collider at CERN with a centre-of-mass energy of at least 100 TeV and with an electron-positron Higgs and electroweak factory as a possible first stage}~\cite{CERN-ESU-015}, thus giving CERN and its international partners the mandate to prepare for a Higgs factory, followed by a future hadron collider with sensitivity to energy scales an order of magnitude higher than those of the high-luminosity LHC (HL-LHC).   

The Future Circular Collider (FCC) programme at CERN features a high-luminosity $\rm e^+ e^-$ electroweak, QCD, flavour, Higgs, and top factory (FCC-ee), followed by a $100$\,TeV pp collider (FCC-hh). It is the most pragmatic, most effective, and safest implementation of a global HEP research infrastructure addressing the above two points at once. The CERN Council endorsed this vision in June 2020. In its June 2021 session, the CERN Council approved~\cite{FCC-FS-plans:202106,FCC-FS-org:202106} and funded~\cite{CERNmtp} the FCC  feasibility study. 

A difficulty in the 2020 strategy process appeared to be that more than one $\rm e^+ e^-$ facility apply as Higgs factory candidates.
With the European Strategy for Particle Physics (ESPP) being updated again at the beginning of 2026, creative alternative proposals are being put forward for CERN in case the preferred plan (FCC-ee) turned out not to be {\it feasible or competitive}: CLIC~\cite{Brunner:2022usy}, for which an implementation plan at CERN~\cite{Aicheler:2652600} had been prepared in 2018, and ILC~\cite{ILCInternationalDevelopmentTeam:2022izu}, for which no implementation plan at CERN exists but for which a {\it vision}~\cite{LCVsion} is being developed. It is the purpose of this note to give the reader factual elements of comparison between FCC-ee, CLIC and ILC@CERN, when acting as Higgs factories.

The unique precision study of a number of Higgs boson properties is indeed a guaranteed deliverable of these facilities, and provides a common platform for comparison. The potential of FCC-ee, CLIC and ILC for Higgs boson physics has been intensely studied and compared, both for the 2018--2020 European Strategy Update (ESU2020)~\cite{deBlas:2019rxi} and for the US Community Study on the Future of Particle Physics (Snowmass 2021)~\cite{deBlas:2022ofj}. The analysis presented in this note is based on the most recent Higgs sensitivities taken from the latest update (July 2024) of Ref.~\cite{deBlas:2022ofj}. 

Specifically, for a given scientific outcome, i.e., for a given set of Higgs coupling precisions, the operation time, the electricity consumption, the greenhouse gas emissions, and the cost are estimated for each of the facilities. With a metric that consists in fixing the desired coupling precisions, the three Higgs factories can be compared on an equal footing, having then the same sensitivity to new physics coupled to the Higgs boson.

The present note is organised as follows. The default operational scenarios currently considered for FCC-ee, CLIC and ILC@CERN are recalled in Section~\ref{sec:scenarios}. In Section~\ref{sec:duration}, the operation time and the electricity consumption of the first two Higgs factory energy stages ($\sqrt{s} = 240$ and 350--365\,GeV for FCC-ee, 380 and 1500\,GeV for CLIC, 250 and 350--500\,GeV for ILC@CERN), when required to reach the same precision as FCC-ee, are evaluated.
An estimate of the corresponding cost and carbon footprint, including the construction and the operation of the facilities, is given in Section~\ref{sec:overview}.
The results are discussed in Section~\ref{sec:discussion}, with in particular a comparison to the findings of earlier analyses~\cite{Janot:2022jtn,Breidenbach:2023nxd,Bloom:2024ixe}. 
Finally, concluding remarks are given in Section~\ref{sec:conclusion}.

\section{Default operational scenarios}
\label{sec:scenarios}

The currently assumed centre-of-mass energies, integrated luminosities, colliding beam longitudinal polarisations, and run durations for FCC-ee, CLIC, ILC@CERN are displayed in Table~\ref{tab:lumipol}. Because, in this note, the Higgs coupling precisions are used to define the comparison metric between the three facilities, only the Higgs factory energy stages are considered. For FCC-ee, about a day of data taking at the Z pole (a few $10^9$ Z's) is needed as well for the Z parameter precisions not to limit the Higgs coupling precision. Four years of running at the Z pole and two years of running at the WW production threshold are also planned, with unique and essential physics results. Including these runs in the comparison (with a metric adapted to their considerable scientific outcome) is beyond the scope of this note. 
Notwithstanding, the corresponding duration, electricity consumption, cost and carbon footprint are added for illustration in the plots of Sections~\ref{sec:duration} and~\ref{sec:overview}.
\begin{table}[ht]
\renewcommand{\arraystretch}{1.25}
\centering
\caption{\label{tab:lumipol} \small Scenarios for future collider options considered in this note for the measurement of Higgs properties. The changes with respect to Ref.~\cite{deBlas:2022ofj} are highlighted in bold and explained in the text. 
\vspace{0.0cm}}
\begin{tabular}{c|c|c|c|c|c}
\hline \begin{tabular}{c} Collider \end{tabular} &  \begin{tabular}{c} Longitudinal \\ Polarisation \\ ($\rm e^-$, $\rm e^+$) (\%)  \end{tabular} & $\sqrt{s}$ (GeV) & \begin{tabular}{c} Integrated \\ Luminosity \\ ($\rm ab^{-1}$) \end{tabular} & 
{\begin{tabular}{c} Time \\ (Years) \end{tabular}}& Ref.  \\ \hline
\begin{tabular}{c}
FCC-ee
\end{tabular} & 
\begin{tabular}{c}
$0,0$
\end{tabular} & 
\begin{tabular}{c}
240 \\ 350 \\ 365
\end{tabular} & 
\begin{tabular}{c}
{\bf 10.8} \\ {\bf 0.42} \\ {\bf 2.70}
\end{tabular} & 
{\begin{tabular}{c}
3 \\ 1 \\ 4
\end{tabular}} & \begin{tabular}{c}
\cite{janot_2024_nfs96-89q08}
\end{tabular}\\ \hline 

\begin{tabular}{c}
CLIC
\end{tabular} & 
\begin{tabular}{c}
 $\pm 80, 0$
\end{tabular} & 
\begin{tabular}{c}
380 \\ 1500 
\end{tabular} & 
\begin{tabular}{c}
{\bf 1.5} \\ 2.5 
\end{tabular} & 
{\begin{tabular}{c}
8 \\ 7 
\end{tabular}} &  
\begin{tabular}{c}
\cite{Brunner:2022usy}
\end{tabular} \\ \hline 

\begin{tabular}{c}
ILC
\end{tabular} & 
\begin{tabular}{c}
 \\ $\pm 80, \pm 30$ \\ \\ $\pm 80, \pm 20$
\end{tabular} & 
\begin{tabular}{c}
250 \\ 350 \\ 500 \\ 1000
\end{tabular} & 
\begin{tabular}{c}
2 \\ 0.1 \\ 4 \\ 8
\end{tabular} & 
\begin{tabular}{c}
{\bf 15}\\ {\bf 1.5}\\ {\bf 11.5} \\ {\bf 13}
\end{tabular} & 
\begin{tabular}{c}
\cite{ILCInternationalDevelopmentTeam:2022izu}
\end{tabular} \\ \hline 

\end{tabular} 
\end{table}

The main differences with respect to the integrated luminosities used in Ref.~\cite{deBlas:2022ofj} come from {\it (i)} the strong recommendation from the FCC mid-term review committees to {\it  ``construct 4 IPs for FCC-ee from the beginning, on the basis of the strong physics case, the improved sustainability in terms of electricity cost per unit of luminosity, and the increased size of the community and diversity of the physics which can be supported.''}, and the further optimisation of the FCC-ee machine parameters in this baseline configuration; {\it (ii)} the optimisation of the CLIC design for the initial stage at 380\,GeV; and {\it(iii)} and the recent decision to abandon the CLIC stage at 3\,TeV~\cite{Steinar}. The FCC-ee and CLIC projections for the Higgs coupling precisions of Ref.~\cite{deBlas:2022ofj} are rescaled here to take these differences into account.

Another important change concerns the energy stage duration for ILC, if operated at CERN. The integrated luminosities and times used in Ref.~\cite{deBlas:2022ofj} were obtained under the assumption that a full calendar year represents eight months running at an efficiency of 75\%~\cite{Bambade:2019fyw}, i.e., $1.6 \times 10^7$ seconds of integrated running. At CERN, realistic years are significantly shorter, as explained in Ref.~\cite{Bordry:2018gri}: this is an important optimisation of resources, as it allows the reduction of both the cost and carbon footprint of each MWh, by running in periods during which  electricity production is abundant and renewable. With the canonical 185 days of operations and 75\% of operational efficiency, $1.2 \times 10^7$ seconds are available every year for collisions, as has always been assumed for FCC-ee and CLIC. The durations of the planned 250, 350+500 and 1000\,GeV runs of ILC@CERN are thus rescaled to 15, 13 and 13 years, instead of the previously assumed 11.5 (of which one year of shutdown for a luminosity upgrade), 9.5 and 10 years. 

\section{Duration and electricity consumption}
\label{sec:duration}

\subsection{Initial energy stages}
\label{sec:initial}

In this section, the durations of the first Higgs factory stages, FCC-ee$_{240}$, CLIC$_{380}$ and ILC$_{250}$, are rescaled to reach the same precision for effective Higgs couplings to fermions and to gauge bosons. With the default durations of Table~\ref{tab:lumipol}, the expected precision on the Higgs couplings to b, c, $\tau$, Z and W after a combination with HL-LHC, as inferred from Ref.~\cite{deBlas:2022ofj}, are shown in Table~\ref{tab:SMEFT250}. Other couplings (such as couplings to gluons, photons, muons, Z$\gamma$, top quarks, etc.) are not considered here because the larger influence of HL-LHC in the combined fit would invalidate the assumption (made throughout) that their precision improves as fast as the square root of the $\rm e^+e^-$ collider integrated luminosity. A stand-alone fit (without the combination with HL-LHC) would be needed for these couplings to be safely included in the procedure.

\begin{table}[h]
\renewcommand{\arraystretch}{1.25}
\centering
\caption{\label{tab:SMEFT250} \small Precision reach (in percentage) on effective couplings from a SMEFT global fit of the Higgs measurements in the first stage of FCC-ee (3 years), CLIC (8 years) and ILC (15 years). The results from the free-$\Gamma_{\rm H}$ fit, scaled from Ref.~\cite{deBlas:2022ofj}, are shown.}
    \begin{tabular}{c|c|c|c}
    \hline 
      \begin{tabular}{l} Precision (\%) on \\  Higgs coupling to \end{tabular} &  \begin{tabular}{l} FCC-ee$_{240}$\end{tabular} &  \begin{tabular}{l} CLIC$_{380}$ \end{tabular} & \begin{tabular}{l} ILC$_{250}$ \end{tabular}  \\ \hline
         b & 0.45  & 0.90 & 0.83 \\ \hline
         c & 0.95 & 3.51 & 1.8 \\ \hline
         $\tau$  & 0.46 & 1.14 & 0.87\\ \hline
         Z & 0.21 & 0.46 & 0.37 \\ \hline
         W & 0.21 & 0.46 & 0.37 \\ \hline
    \end{tabular}
\end{table}

Because of the 5 to 7 times larger integrated luminosity, the FCC-ee precision is typically twice better than that of CLIC and ILC. The significant benefits from the incoming beam longitudinal polarisation naturally available at linear colliders do not suffice to compensate for the much larger event rates at FCC-ee. More importantly, the time needed for FCC-ee to reach the precision listed in Table~\ref{tab:SMEFT250} is only three years, while it would take eight and fifteen years for CLIC and ILC as shown in Table~\ref{tab:timeenergy250}. 
\begin{table}[h]
\renewcommand{\arraystretch}{1.25}
\centering
\caption{\label{tab:timeenergy250} \small Time needed, yearly electricity consumption, and total electricity consumption by the planned first stages of FCC-ee, CLIC and ILC@CERN, to reach the precision listed in Table~\ref{tab:SMEFT250}. The two numbers for the yearly ILC electricity consumption correspond to the first and second half of the run, before/after the luminosity upgrade.}
    \begin{tabular}{c|c|c|c}
    \hline 
     Collider &  FCC-ee$_{240}$ &  CLIC$_{380}$ & ILC$_{250}$ \\ \hline
         Duration (years) & 3  & 8 & 15 \\ \hline
         Yearly electricity consumption (TWh) & 1.33 & 0.6 & 0.6--0.7 \\ \hline
         Total electricity consumption (TWh) & 4  & 5 & 9\\ \hline
    \end{tabular}
\end{table}

The yearly electricity consumptions are documented in Ref.~\cite{Brunner:2022usy} for CLIC$_{380}$ and in the Feasibility Study Mid-Term report~\cite{auchmann_2024_511pr-rd590} for FCC-ee$_{240}$: they  amount to 0.6\,TWh and 1.33\,TWh respectively. The yearly ILC$_{250}$ electricity consumption at CERN is not documented in Ref.~\cite{ILCInternationalDevelopmentTeam:2022izu}, but is inferred here from the CLIC energy consumption of 0.6\,TWh, augmented by the ILC-to-CLIC electrical power (111--128\,MW for ILC$_{250}$ and 110\,MW for CLIC$_{380}$) ratio. 

In short, FCC-ee reaches twice better precision with up to 2.5 times smaller electricity consumption and up to 5 times quicker. These multiple advantages are summarised by evaluating the time (in years) and the energy consumption (in TWh) needed for CLIC and ILC@CERN to reach the same coupling precision as obtained by FCC-ee in three years. To do so, the total integrated luminosity in the CLIC (ILC) first stage is increased by the square of the CLIC(ILC)-to-FCC-ee precision ratio  (Table~\ref{tab:SMEFT250}), and the number of additional years needed to reach these new luminosities is inferred from the design yearly integrated luminosity of CLIC$_{380}$ (0.276\,ab$^{-1}$/year or $2.3\times 10^{34}$\,cm$^{-2}$s$^{-1}$) and ILC$_{250}$ (0.324\,ab$^{-1}$/year or $2.7\times 10^{34}$\,cm$^{-2}$s$^{-1}$ after the luminosity upgrade). 

\begin{figure}[htbp]
\centering
\includegraphics[width=1.0\textwidth]{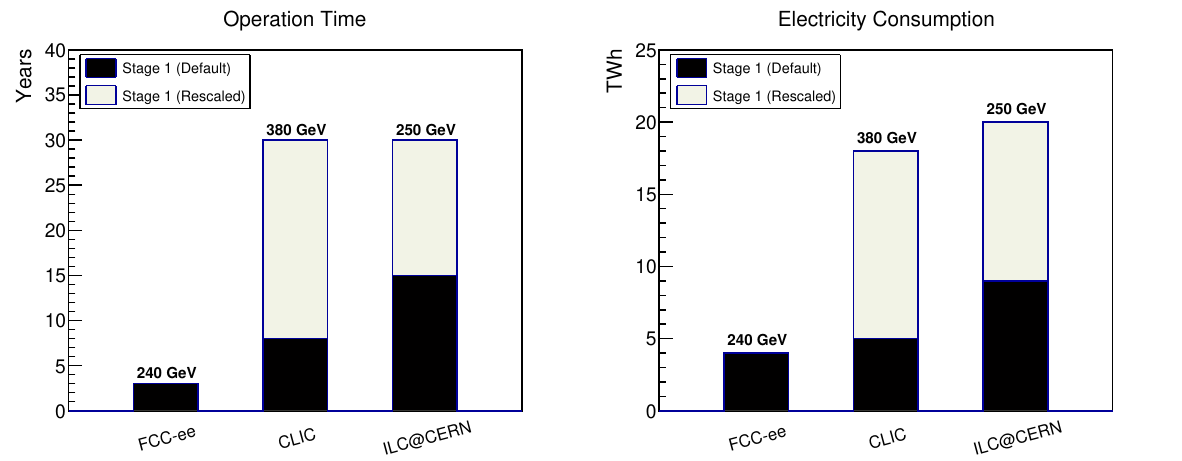}
\caption{\label{fig:time250} \small (Left panel) Operation time [years] and (right panel) electrical energy consumption [TWh], for the first stage of three Higgs factory options at CERN (FCC-ee, $\sqrt{s}=240$\,GeV; CLIC, $\sqrt{s} = 380$\,GeV; and ILC@CERN,$\sqrt{s}=250$\,GeV), in the default scenario (black) and rescaled to reach the same Higgs coupling precision as FCC-ee (off-white). 
}
\end{figure}

The result of this exercise, summarised in Table~\ref{tab:time250} and displayed in Fig.~\ref{fig:time250}, shows that about 30 years are needed for ILC@CERN and CLIC during their initial stages to reach the precision of the three-years run of FCC-ee at 240\,GeV, almost independently of the coupling considered.\footnote{With a similar reasoning, FCC-ee would need less than a year to reach the precision of the default 15-years ILC run at 250\,GeV or of the 8-years CLIC run at 380\,GeV.} The linear collider electricity consumption during these 30 years of operation would then be about five times larger than at FCC-ee.\footnote{For ILC@CERN, these 30 years would allow an integrated luminosity of 6.5\,ab$^{-1}$ to be accumulated, to be compared to the 10.8\,ab$^{-1}$ of FCC-ee: longitudinal beam polarisation is therefore equivalent to a 66\% increase in luminosity. The possibility of injecting longitudinally polarised beams at FCC is being considered, but is not expected to provide physics results that cannot be achieved without.}

A similar conclusion had already been reached in Ref.~\cite{Janot:2022jtn}.

\begin{table}[h]
\renewcommand{\arraystretch}{1.25}
\centering
\caption{\label{tab:time250} \small Time needed for CLIC$_{380}$  and for ILC$_{250}$@CERN to deliver the integrated luminosity needed to reach the same precision as FCC-ee$_{240}$ in three years, for selected couplings to fermions and gauge bosons. The last row indicates the total electricity consumption for the average 30 years of operation. (The CLIC duration for the c coupling seems off, probably because of rounding errors after/before the fit in Ref.~\cite{deBlas:2022ofj}, and is not included in the average.)}
    \begin{tabular}{c|c|c|c}
    \hline 
     Duration (years) &  FCC-ee$_{240}$ &  CLIC$_{380}$ & ILC$_{250}$ \\ \hline
         b & 3 & 24  & 30 \\ \hline
         c & 3 & 77  & 31 \\ \hline
    $\tau$ & 3 & 36  & 31\\ \hline
         Z & 3  & 29  &  28    \\ \hline
         W & 3  &  29  &  28   \\ \hline \hline
  Average duration (years) & 3  & \multicolumn{2}{c}{30}\\ \hline
  Electricity consumption (TWh) & 4  & 18 & 20  \\ \hline    \end{tabular}
\end{table}

\subsection{Energy upgrades}
\label{sec:upgrades}
All facilities do include higher energy upgrades, which benefit the Higgs coupling precisions.
It is assumed in the rest of this note that these upgrades are affordable and approved. 
The baseline operation model of FCC-ee includes a run at the t\=t threshold and slightly above, up to $\sqrt{s} = 365$\,GeV. Energy upgrades at 350--500\,GeV for ILC and at 1.5\,TeV for CLIC are also considered. The precision reach of the three colliders after these second stages with the default integrated luminosities of Table~\ref{tab:lumipol}, as inferred from Ref.~\cite{deBlas:2022ofj}, is displayed in Table~\ref{tab:SMEFT500} for the same couplings as in Table~\ref{tab:SMEFT250}. These second stages also allow the Higgs self-coupling to be probed. A specific discussion can be found in Section~\ref{sec:self}.

\begin{table}[h]
\renewcommand{\arraystretch}{1.25}
\centering
\caption{\label{tab:SMEFT500} \small Precision reach (in percentage) on effective couplings from a SMEFT global fit of the Higgs measurements after the planned second stages of FCC-ee (365\,GeV), CLIC (1.5\,TeV) and ILC (500\,GeV), i.e., after 8, 15 and 28 years of operation, respectively. The results from the free-$\Gamma_{\rm H}$ fit, scaled from Ref.~\cite{deBlas:2022ofj}, are shown.}
    \begin{tabular}{c|c|c|c}
    \hline 
      \begin{tabular}{l} Precision (\%)  \\  on coupling to \end{tabular} & \begin{tabular}{l} FCC-ee$_{240+365}$\end{tabular} 
      & \begin{tabular}{l} CLIC$_{380+1500}$ \end{tabular} 
      & \begin{tabular}{l} ILC$_{250+500}$ \end{tabular} \\ \hline
         b & 0.40 & 0.56 & 0.56  \\ \hline
         c & 0.89 & 1.81 & 1.2 \\ \hline
         $\tau$ & 0.42 & 0.89 & 0.63 \\ \hline
         Z & 0.17 & 0.36 & 0.26 \\ \hline
         W & 0.17 & 0.37 & 0.26 \\ \hline
    \end{tabular}
\end{table}

The additional time needed and electricity consumption for these second stages are indicated in Table~\ref{tab:timeenergy500}, together with the total time and electricity consumption for the first two stages altogether.

\begin{table}[h]
\renewcommand{\arraystretch}{1.25}
\centering
\caption{\label{tab:timeenergy500} \small Additional time needed, yearly electricity consumption, and total electricity consumption by the planned second stages of FCC-ee, CLIC and ILC@CERN (with $1.2\times 10^7$\,seconds/year), to reach the precisions listed in Table~\ref{tab:SMEFT500}. The last two rows indicate the total duration and energy consumption during the default first and second stages.}
    \begin{tabular}{c|c|c|c}
    \hline 
     Collider &  FCC-ee$_{240+365}$ &  CLIC$_{380+1500}$ & ILC$_{250+500}$ \\ \hline
         Additional duration (years) & 5 & 7  & 13 \\ \hline
         Yearly energy consumption (TWh) & 1.77  & 1.8 & 1.2\\ \hline
         Additional energy consumption (TWh) & 9  & 13 & 15\\ \hline
        Total duration (years) & 8  & 15 & 28 \\ \hline         Total energy consumption (TWh) & 13  & 17 & 24\\ \hline
    \end{tabular}
\end{table}

As in the previous section, the time (in years) and the energy consumption (in TWh) needed for CLIC and ILC@CERN to reach the same precision as FCC-ee with their first and second stages are evaluated. To do so, the default CLIC (ILC) integrated luminosities in the first and the second stages are thus increased by the square of the CLIC(ILC)-to-FCC-ee precision ratio, and the number of additional years needed to reach these new luminosities is inferred from the design yearly integrated luminosities: 0.276 and 0.444\,ab$^{-1}$ (2.3 and $3.7 \times 10^{34}$\,cm$^{-2}$s$^{-1}$) for CLIC at 380 and 1500\,GeV; and 0.324 and 0.432 \,ab$^{-1}$ (2.7 and $3.6\times 10^{34}$\,cm$^{-2}$s$^{-1}$) for ILC at 250 and 500\,GeV. 

\begin{figure}[htbp]
\centering
\includegraphics[width=1.0\textwidth]{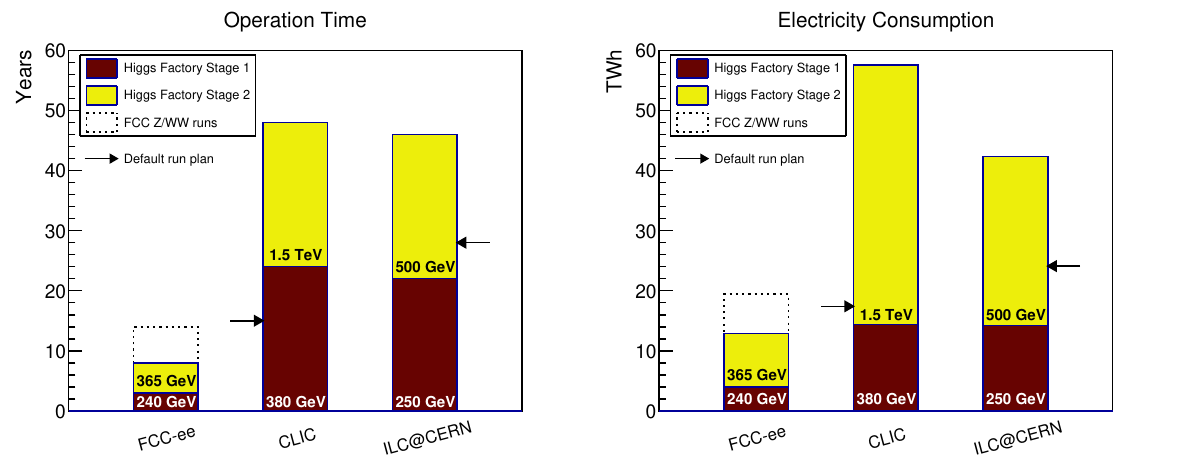}
\caption{\label{fig:time500} \small (Left panel) Operation time in years and (right panel) electrical energy consumption in TWh, for the two stages of three possible Higgs factory options at CERN (ILC@CERN with $\sqrt{s}=250$ and 500\,GeV, CLIC with $\sqrt{s} = 380$ and 1500\,GeV, and FCC-ee with $\sqrt{s}=240$ and 365\,GeV) to reach the FCC-ee Higgs coupling precision. For illustration, the operation time and the electricity consumption of the runs at the Z pole and WW threshold of FCC-ee, not used for the comparison made in this note, are indicated with a dashed line. The operation times and electricity consumptions of the default linear collider run plans, leading to degraded coupling precision, are indicated by arrows.
}
\end{figure}

The conclusion, illustrated in Table~\ref{tab:time500} and Fig.~\ref{fig:time500}, is that about 48 years for CLIC and 46 years for ILC@CERN are needed during their first and second stages to reach the precision of the eight-years FCC-ee run at 240 and 365\,GeV, increased to more than half a century once the regular shutdown periods for maintenance and upgrades are included.\footnote{With a similar reasoning, FCC-ee would need about 4 (2) years to reach the precision of the default 28 (15)-years ILC (CLIC) run at 250/500 (380/1500)\,GeV.} The linear collider electricity consumption during this half a century of operation would then be three to four times larger than that of the FCC-ee run, for the same physics outcome. Even after these second stages a priori favourable to linear colliders, FCC-ee operations therefore remain -- by large factors -- the most sustainable operations of all. The contribution of the facility construction is addressed in the next section.
\vskip -.3cm
\begin{table}[h]
\centering
\renewcommand{\arraystretch}{1.25}
\caption{\label{tab:time500} \small Time needed for CLIC$_{380+1500}$  and for ILC$_{250+500}$ @CERN to deliver the integrated luminosity needed to reach the same precision as FCC-ee$_{240+365}$ in eight years, for selected couplings. The last row indicates the total energy consumption for the average 46 years of operation. (The CLIC duration for the coupling to the b seems off, probably because of rounding errors after/before the fit in Ref.~\cite{deBlas:2022ofj}, but is conservatively included in the average.) 
}
    \begin{tabular}{c|c|c|c}
    \hline 
     Duration (years) &  FCC-ee$_{240+365}$ &  CLIC$_{380+1500}$ & ILC$_{250+500}$ \\ \hline
         b & 8 & 26 & 43 \\ \hline
         c & 8 & 50 & 41 \\ \hline
    $\tau$ & 8 & 54 & 47 \\ \hline
         Z & 8 & 54 & 49 \\ \hline
         W & 8 & 56 & 49 \\ \hline \hline
  Average duration (years) & 8 & 48 & 46 \\ \hline
  Electricity consumption (TWh) & 13  & 55  & 41 \\ \hline    
  \end{tabular}
\end{table}

A third stage at a centre-of-mass energy of 1\,TeV is also envisioned for ILC, for a planned duration of 13 years (after 2 years shutdown) if implemented at CERN, and an electricity consumption of 21\,TWh. In the default operation model, the total duration would exceed 42 years, for a total electricity consumption of 46\,TWh. These figures are similar to those of the extended first two stages in Table~\ref{tab:timeenergy500} (46 years and 41\,TWh), but are yet insufficient to bring the Higgs coupling precision to the level of FCC-ee$_{250+365}$. It would take 60 years and 67\,TWh for ILC$_{250+500+1000}$ to reach the FCC-ee precision. 
For the same reason as the third CLIC stage was recently abandoned, this third stage of ILC is not considered for implementation at CERN in the following discussion. 

\section{Partial estimates of cost and carbon emissions}
\label{sec:overview}

\subsection{Cost}

The overall cost includes the cost of operation and the cost of construction of the facility and has been estimated as follows, for a given physics outcome.
\begin{itemize}
    \item To start with, the operation cost is approximated to the cost of electricity. Predicting this cost in 20 years from now is rather uncertain, because of the likely inflation and the cost of the large investment required for the development and operation of renewable energy sources. Consequently, the current cost of 80 euros per MWh~\cite{Fabiola} is used. The resulting operation cost, shared among the two Higgs factory energy stages, is shown in Table~\ref{tab:operationcost} for the three facilities.

    \begin{table}[h]
\centering
\renewcommand{\arraystretch}{1.25}
\caption{\label{tab:operationcost} \small Cost of electricity for FCC-ee$_{240+365}$, CLIC$_{380+1500}$  and ILC$_{250+500}$@CERN to reach the same precision as FCC-ee$_{240+365}$ for the Higgs couplings to fermions and gauge bosons.
}
    \begin{tabular}{c|c|c|c}
    \hline 
     Cost (Billion euros) &  FCC-ee$_{240+365}$ &  CLIC$_{380+1500}$ & ILC$_{250+500}$ \\ \hline  
     First stage  & 0.3 & 1.2 & 1.1  \\ \hline
     Second stage & 0.7 & 3.4 & 2.3 \\ \hline \hline
     Total        & 1.0 & 4.6 & 3.4 \\ \hline
  \end{tabular}
\end{table}

    \item For the construction (including civil engineering, infrastructures, and the components of the collider), a full FCC cost assessment was performed at the occasion of the FCC Feasibility Study mid-term review, in 2024 and led to 16.1 billion euros (15 billion CHF), out of which 1.6 billion euros is needed for 365\,GeV stage. The cost for the civil engineering and infrastructures re-used for FCC-hh amounts to 8.3 billion euros. An estimate of the CLIC cost, made in 2018, can be found in Ref.~\cite{Brunner:2022usy}: 6.3 billion euros for the drive-beam design of the 380\,GeV stage (increased to 7.8 billion euros for the klystron design), and 5.5 billion euros for the second stage. The drive-beam design is chosen here for display, and a 17\% cost increase is applied to account for the inflation between 2018 and 2024~\cite{flopriv}. No cost estimate exists for ILC@CERN. Here, a {\it guesstimate} proposed in the Snowmass Implementation Report~\cite{Roser:2022sht}, roughly 8.5 and 4.8 billion euros for the two energy stages, is taken for display. Updated, likely higher, cost estimates are expected for the next ESU. 

\end{itemize}

\begin{figure}[h]
\centering
\includegraphics[width=1.0\textwidth]{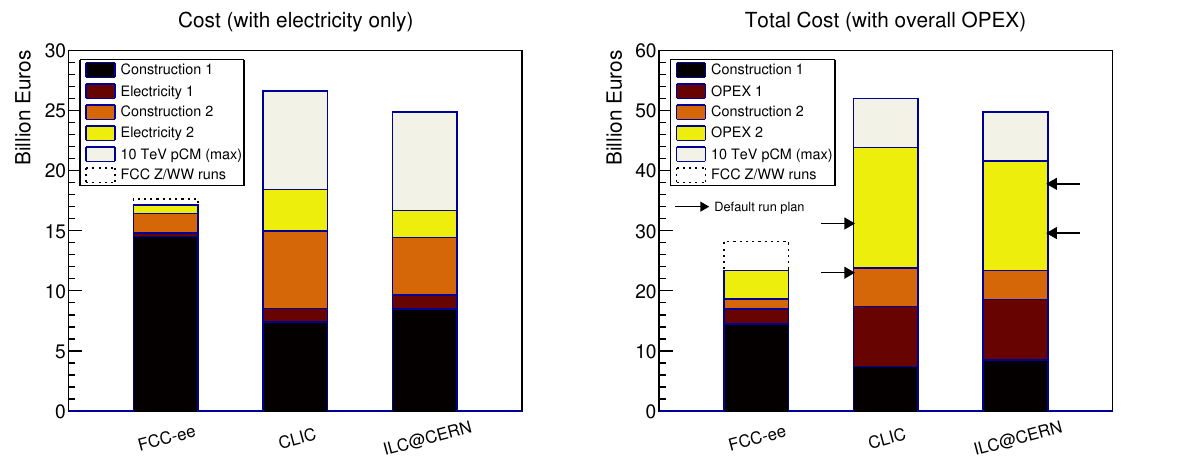}
\caption{\label{fig:Cost} \small Estimated cost in billion euros, for the construction and operation of the two energy stages (labelled 1 and 2) of three possible Higgs factory options at CERN (FCC-ee with $\sqrt{s}=240$ and 365\,GeV, CLIC with $\sqrt{s} = 380$ and 1500\,GeV and ILC@CERN with $\sqrt{s}=250$ and 500\,GeV) to reach the FCC-ee Higgs coupling precision.   
(Left) Only the electricity cost in included in the cost of operations; (Right) The cost of staff, contractors, spares and maintenance during operation is included as well.
For illustration, the cost of the runs at the Z pole and WW threshold of FCC-ee, and the maximal cost for the tunnel(s) and infrastructure of a 10\,TeV pCM collider after CLIC or ILC@CERN, are indicated on top. The estimated ranges for the cost of the default linear collider run plans (with or without the 10\,TeV pCM collider tunnel), leading to degraded coupling precision, are indicated by arrows.}
\vskip -0.5cm
\end{figure}

The different contributions (construction and electricity consumption of the Higgs factory two stages) are displayed in left panel of Fig.~\ref{fig:Cost} for the three options. Operating costs due to electricity consumption, however, are only a minimal part of the overall operating cost. Historically, it is evaluated that the cost of staff, contractors, spares, maintenance and operation (so called ``OPEX'') amounts to $\sim 5$\% of the total construction cost (tunnel, infrastructure, components) every year~\cite{flopriv} for CERN colliders. This is illustrated in the right panel of Fig.~\ref{fig:Cost}. In addition, the vision of the field, as stated by Snowmass, the P5 panel, and also ESU2020, is to build a ``10 TeV parton centre-of-mass energy (10 TeV pCM) collider'' after the Higgs factory, the construction of which would add cost to the linear collider paths. For the circular collider option, the FCC-hh civil engineering and infrastructure cost is already included in FCC-ee. This cost is taken as a proxy for the maximal cost for the tunnel(s) and infrastructure of the 10 TeV pCM that would follow a linear Higgs factory at CERN, and added for illustration in Fig.~\ref{fig:Cost}.

\vskip -0.1cm

\subsection{Carbon budget}

A comparison of the carbon footprint expected for the three facilities, for a given physics outcome, can be made with today's figures. A projection can also be made at the time of construction (2033-2040) and operation (after 2045) of the chosen Higgs factory. 

\begin{itemize}
\item As for the cost, the carbon emissions resulting from the operation of the collider is limited to the carbon intensity of the electricity production in France (as CERN buys its electricity for ``Electricité De France, EDF''). Today, this carbon intensity amounts to 56\,kg CO$_2$e per MWh. For the future, it can be estimated with a power purchase scenario based on the currently communicated ADEME intensity figures~\cite{ademe}. In this approach, we assumed a 80\% to 90\% coverage from renewable energy sources by the start of the Higgs factory operations. Based on the currently communicated figures from renewable sources (in kg CO$_2$e per MWh, hydro-electric: 6.0; photo-voltaic: 25.2; wind onshore: 14.1; wind offshore: 15.6) and non-renewable sources (nuclear: 3.7; other sources: 52), this mix gives a carbon intensity between 15 and 25 kg CO$_2$e per MWh. A value of 20 kg CO$_2$e per MWh is chosen here for display: 0.28 and 0.56 Megatons for ILC@CERN, 0.29 and 0.86 Megatons for CLIC, and 0.08 and 0.18 Megatons for FCC-ee, for the two energy stages in sequence. 
\vskip 0.05cm
\item Life cycle assessments of the carbon budget exist for ILC and CLIC~\cite{footprintLC} as well as for FCC-ee~\cite{johannes}. For the civil engineering and infrastructures 
a generic total carbon footprint of 0.266\,Mt CO$_2$e is inferred for ILC$_{250}$ (length 20.5\,km), and was scaled here to 0.402\,Mt for ILC$_{500}$ (length 31\,km). The same report gives 0.127\,Mt CO$_2$e for CLIC$_{380}$ in its drive-beam design, increased to 0.296\,Mt CO$_2$e for CLIC$_{1500}$. In the CLIC first-stage klystron design, the total construction carbon footprint would be 0.496\,Mt CO$_2$e instead. The former value is taken here. A reduction of about 40\% is predicted by the time of construction. For the FCC-ee tunnel and infrastructures, the generic carbon footprint of 1.183\,MtCO$_2$e is 
reduced to 0.526\,Mt when based on Environmental Product Declarations (EPDs) of materials that are available on the market today. This figure may be further reduced by the time of construction but, conservatively, no reduction factor is applied here for FCC-ee.
\end{itemize}
\vskip -0.1cm
\begin{figure}[h]
\centering
\includegraphics[width=1.0\textwidth]{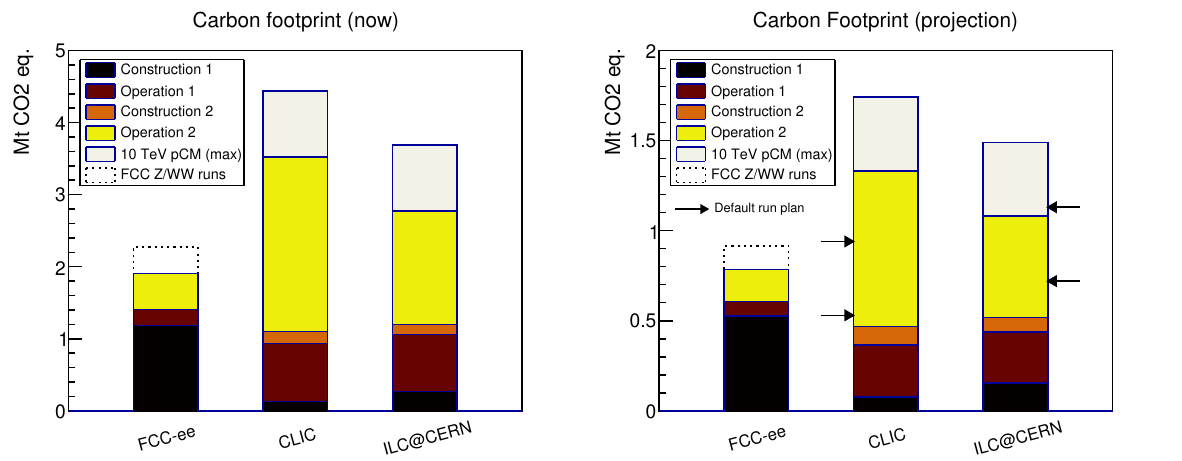}
\caption{\label{fig:Sustainability} \small Estimated carbon footprint in Mt CO$_2$e estimated with today's figures (left panel) and projected by the time of construction and operation (right panel), for the two energy stages (labelled 1 and 2) of three Higgs factory options at CERN (FCC-ee with $\sqrt{s}=240$ and 365\,GeV, CLIC with $\sqrt{s} = 380$ and 1500\,GeV, and ILC@CERN with $\sqrt{s}=250$ and 500\,GeV) to reach the FCC-ee Higgs coupling precision. 
For illustration, the carbon emissions of the runs at the Z pole and WW threshold of FCC-ee, and the maximal budget for the tunnel(s) and infrastructure of a 10\,TeV pCM collider after CLIC or ILC@CERN, are indicated on top. The carbon budget for the accelerator components is not included. The estimated ranges for the carbon budget of the default linear collider run plans (with or without the 10\,TeV pCM collider tunnel), leading to degraded coupling precision, are indicated by arrows.
\vskip -0.5cm
}
\end{figure}

The different contributions are displayed in Fig.~\ref{fig:Sustainability}
for the three collider options, with today's figures and for projection by the time of construction and operation. The FCC-ee tunnel would then be reused with no additional emissions for FCC-hh. After a linear collider, civil engineering and infrastructures would be needed for the 10 TeV pCM collider. The FCC-hh tunnel carbon footprint is taken as a proxy for the maximal carbon emissions, which adds 0.918 (0.408)\,Mt CO$_2$e, a little less than for the FCC-ee tunnel, as no enlargement around the 4 IPs would be required in that case. The carbon budget for the accelerator components, currently under evaluation, is not included.

\subsection{Summary}
\begin{figure}[htbp]
\centering
\includegraphics[width=1.0\textwidth]{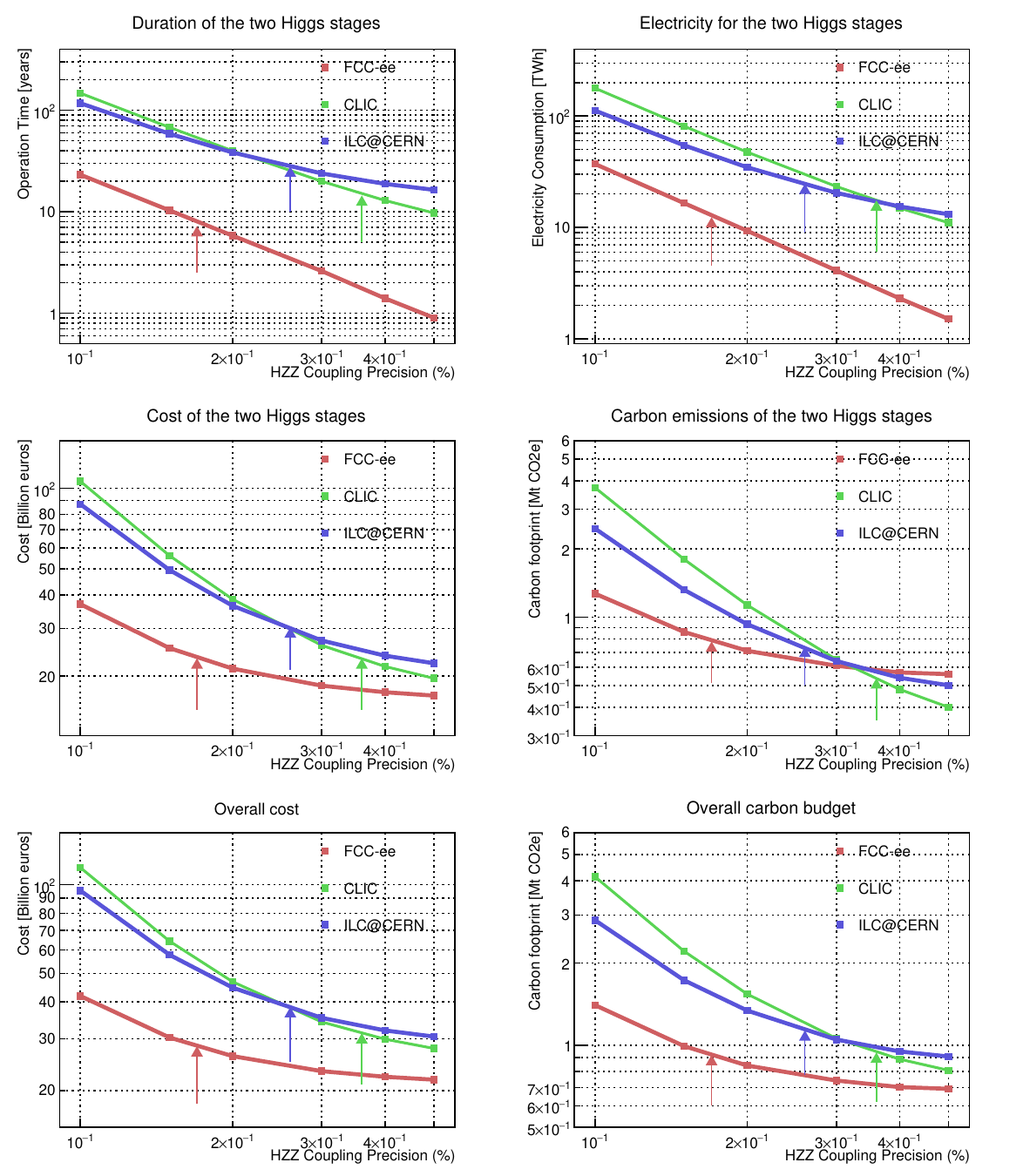}
\caption{\label{fig:precision} \small From left to right, and from top to down: Operation time in years, electricity consumption in TWh, estimated cost in billion euros, and estimated carbon emissions in Mt CO$_2$e, for the construction and operation of the two stages of three Higgs factory options at CERN (FCC-ee with $\sqrt{s}=240$/365\,GeV in red, CLIC with $\sqrt{s} = 380$/1500\,GeV in green, and ILC@CERN with $\sqrt{s}=250$/500\,GeV in blue) as function of the desired precision on the Higgs coupling to the Z boson (as obtained from the SMEFT fit of Ref.~\cite{deBlas:2022ofj}). For illustration, the overall cost and carbon budget are shown in the two bottom plots, with the runs at the Z pole and WW threshold of FCC-ee, and with the tunnel(s) and infrastructure of a 10 TeV pCM after CLIC or ILC@CERN. The three arrows in the each plot indicate the default run plans of FCC-ee, CLIC and ILC@CERN. For example, FCC-ee would need only 2 (4) years to reach the HZZ coupling precision that CLIC (ILC@CERN) would achieve in 15 (28) years.}

\end{figure}

In short, and as illustrated in Figs.~\ref{fig:time500},~\ref{fig:Cost}, \ref{fig:Sustainability} and~\ref{fig:precision}, FCC-ee is -- for the precision on Higgs couplings to gauge bosons and fermions reached by FCC-ee in eight years of operation -- the least disruptive Higgs factory option for CERN, in terms of operation time (six times faster), electricity consumption (four-to-five times less greedy), environmental impact (twice milder), and overall cost (almost twice more cost effective). 
The dependence of this observation as a function of the desired precision on the Higgs coupling to the Z boson is shown in Fig.~\ref{fig:precision}. Not only is the FCC-ee the least disruptive option for CERN for any value of the desired Higgs coupling precision, but it also allows the best precision to be reached in the shortest time, with the smallest electricity consumption, with the best cost effectiveness, and with competitive carbon budget, when the default run plans are compared to each other.

\section{Discussion}
\label{sec:discussion}

\subsection{The Higgs self-coupling}
\label{sec:self}

It is often argued that a 500\,GeV run at ILC and a 1.5\,TeV run at CLIC are needed to have a first measurement of the triple Higgs self-coupling with Higgs pair production, in addition to the couplings to gauge bosons and fermions. With the currently planned runs at these energies, a modest precision of 36\% (27\%)\footnote{Improvements are envisioned~\cite{Bliewert:2024ftv} on these precisions.} can be achieved at CLIC (ILC) from the analysis of Higgs pair production~\cite{deBlas:2019rxi,DiMicco:2019ngk}.
With the extended run needed to reach the FCC-ee precision on the other couplings, both ILC$_{500}$@CERN and CLIC$_{1500}$ would improve this figure to 20\% after 50 years of operation.

\begin{figure}[htbp]
\centering
\includegraphics[width=0.32\textwidth]{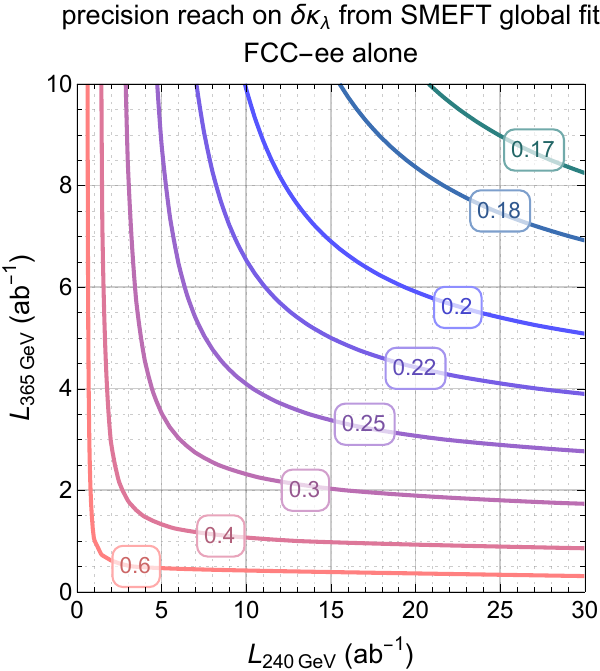}
\includegraphics[width=0.32\textwidth]{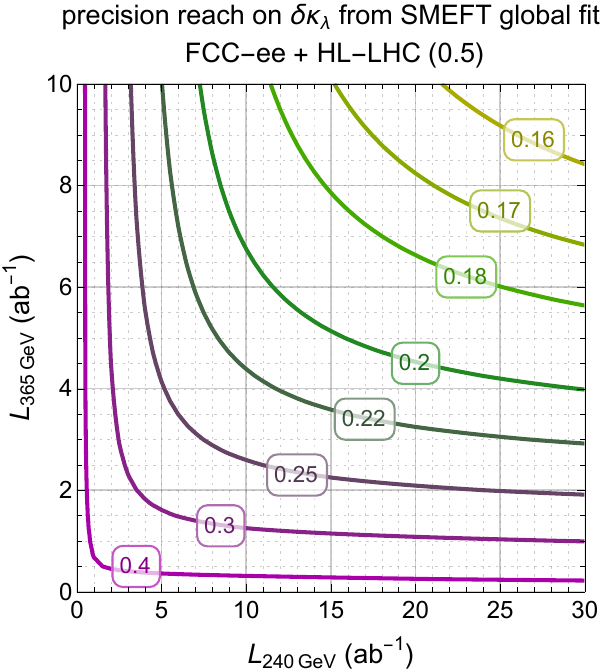}
\includegraphics[width=0.32\textwidth]{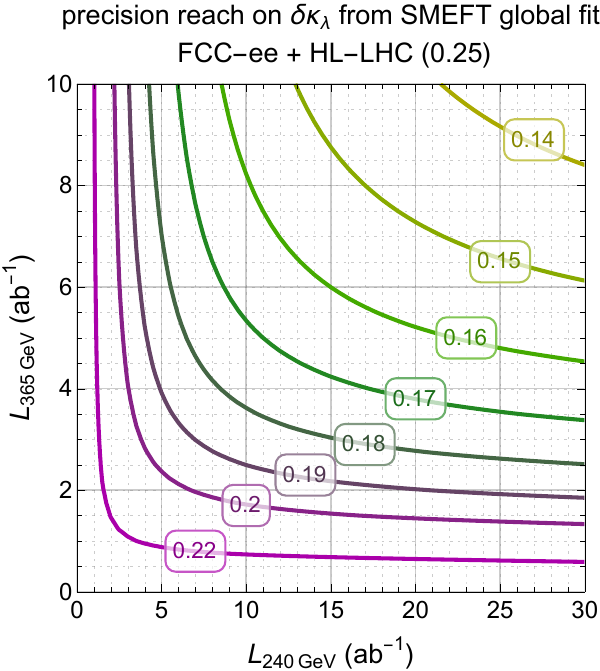}

\caption{\label{fig:kappalambda} \small Precision on the Higgs self-coupling as a function of the FCC-ee integrated luminosities at 240 and 365\,GeV. (Left) FCC-ee alone; (Middle) FCC-ee combined with a 50\% precision at HL-LHC; (Right) FCC-ee combined with a 25\% at HL-LHC. (Courtesy: Jiayin Gu)}

\end{figure}

On the one hand, such a precision is not decisively better than what HL-LHC would achieve\footnote{There is good hope, based on the progress made recently with modern analysis techniques, on the study of additional decay channels, and on the improved capabilities of the ATLAS and CMS detector upgrades, that the HL-LHC would deliver a precision of around 25 to 30\% on the Higgs self-coupling measurement through the analysis of Higgs pair production~\cite{marupriv}.} with Higgs pair production 50 years before. (A similar reasoning applies to the measurement of the top-quark Yukawa coupling.) A measurement of the Higgs self-coupling with a similar precision (27\%) can also be made by FCC-ee alone via its effect on the single Higgs production cross sections at 240 and 365\,GeV (left panel of Fig.~\ref{fig:kappalambda}), with integrated luminosities of 10.8 and 3.12 ab$^{-1}$, respectively. A combination with the 50\% (25\%) HL-LHC precision would improve the stand-alone FCC-ee precision to 24\% (18\%), as shown in the middle and right panels of Fig.~\ref{fig:kappalambda}, which also show how this precision would benefit from higher FCC-ee luminosities. 

On the other hand, and much more importantly, during the estimated 50 years of extended ILC/CLIC operations at just 250-500/380-1500\,GeV, the full FCC integrated programme would be completed, yielding in particular a 3.4\% precision measurement\footnote{And probably significantly better, with the improvement prospects pioneered for HL-LHC} of the Higgs self-coupling from the 100\,TeV hadron collider~\cite{Mangano:2020sao}, hence with a much improved sensitivity to new physics modifying the Higgs potential~\cite{DiMicco:2019ngk,Durieux:2022hbu}. The FCC-hh is also and primarily a discovery machine that is ultimately required to search for new physics and answer some of the fundamental questions in particle physics within a generation. The breadth of couplings and new phenomena that can be explored in these 50 years with the FCC integrated programme, most of which being simply inaccessible to linear colliders, is a paramount consideration in planning for the future. It would be a scientific and strategic mistake to limit the future ambitions of particle physics to the measurement of the Higgs self-coupling, with a modest two-digit precision figure after 50 years of operation.

\subsection{Other (unique) Higgs opportunities at FCC}

The FCC-ee advantage in operation time is striking for the precise measurements of already demonstrated Higgs decays (b\=b, $\tau^+\tau^-$, gg, ZZ, WW) and for $\rm H \to c\bar c$. Some couplings are not yet demonstrated such as the coupling to the s quark or the electron Yukawa coupling. The FCC-ee is the best/only place to attempt their measurement in a reasonable amount of time, with a sensitivity close to the Standard Model prediction, thanks to its large luminosity and its four interaction points. For the electron Yukawa coupling, the ability to produce the Higgs boson directly at $\sqrt{s} = m_{\rm H}$ with reduced centre-of-mass energy spread and excellent centre-of-mass energy calibration with resonant depolarisation are game changers, unique to circular colliders. 

Other couplings, such as couplings to $\gamma\gamma$, $\rm Z\gamma$, or $\mu^+\mu^-$, or the decay to invisible final states, will need a much larger number of Higgs bosons  than is conceivable to produce at $\rm e^+e^-$ colliders, because of the limited corresponding branching fractions. \begin{table}[htbp]
\renewcommand{\arraystretch}{1.25}
\centering
\caption{\small Precision (in \%) on Higgs boson couplings from Ref.~\protect\cite{deBlas:2019rxi}, in the $\kappa$ framework without (first numbers) and with (numbers between brackets, when available) HL-LHC projections, after a couple years at FCC-hh, and after completion of the full ILC$_{1\,{\rm TeV}}$ and CLIC$_{3\,{\rm TeV}}$ programmes. Also indicated are the 95\% C.L. sensitivity on the ``invisible'' Higgs branching fraction.
\label{tab:FCChh}} 
\vspace{2mm}
\begin{tabular}{c|c|c|c}
\hline Coupling precision (\%) & ILC$_{0.25+0.5+1\,{\rm TeV}}$ & {CLIC$_{0.38+1.5+3\,{\rm TeV}}$}  & FCC (ee+hh) \\ \hline
$g_{\rm H\mu\mu}$ (\%) & 6.3 (3.6)  & 5.9 (3.5) & 0.43 (0.43) \\ 
$g_{\rm H\gamma\gamma}$ (\%) & 1.9 (1.1) & 2.3 (1.1) & 0.32 (0.32) \\ 
$g_{\rm HZ\gamma}$ (\%) & -- (10.) & 7. (5.7) & 0.71 (0.70)\\ 
$g_{\rm Htt}$ (\%) & 1.6 (1.4) & 2.7 (2.1) &  1.0 (0.95) \\ 
$g_{\rm Hgg}$ (\%) & 0.67 (0.63) & 0.96 (0.86)  & 0.52 (0.50)\\ 
$g_{\rm HHH}$ (\%) & 10. & 9. & 3.4 \\ \hline
BR$_{\rm inv}$ (\%) & 0.22 & 0.61 & 0.024 \\ \hline
\end{tabular} 
\end{table}
These couplings will be better measured at FCC-hh, where several billions Higgs bosons are expected to be produced. The coupling to gluons, which affect the production of Higgs boson at hadron colliders, will also be significantly better measured at FCC-hh. The achievable FCC-hh precisions, in combination with FCC-ee, are listed in Table~\ref{tab:FCChh} and compared to what could be achieved with the full linear collider programmes (including the highest centre-of-mass energy stages), as obtained Tables 5 and 29 of Ref.~\cite{deBlas:2019rxi}.

\subsection{Operation time optimisation}
\label{sec:optimization}

In Section~\ref{sec:upgrades}, no attempt was made to optimise the respective durations of the two energy stages. The total integrated luminosity is simply varied to reach a given coupling precision, and the integrated luminosity ratio between the two stages -- supposedly optimised with respect to the collider scientific outcome by the respective teams -- is kept constant. If the optimisation were to be done with respect to overall duration/cost/carbon budget of the Higgs factory run instead, for a given precision on the Higgs boson couplings to fermion and gauge bosons, it would be preferable to always stay at the lowest centre-of-mass energy (240\,GeV for FCC-ee, 250\,GeV for ILC@CERN, and 380\,GeV for CLIC). 

Five years would be enough at 240\,GeV for FCC-ee to reach (or exceed) the precisions of the planned 240+365\,GeV runs, while about 50 years would still be needed for CLIC and ILC@CERN to achieve the same precisions, with reduced electricity consumption, cost, and carbon emissions, as shown in Table~\ref{tab:optimisation}. Surely, however, the members of the linear and circular Higgs factory efforts would be keen on putting forward that, should the second energy stage be abandoned, considerable physics opportunities might be lost. 

\begin{table}[h]
\renewcommand{\arraystretch}{1.25}
\centering
\caption{\label{tab:optimisation} \small Operation time, electricity consumption, estimated construction and operation cost and carbon budget, for FCC-ee$_{240}$, CLIC$_{380}$ and ILC@CERN$_{250}$ to achieve the same Higgs coupling precision as the planned FCC-ee$_{240+365}$ runs. The last two rows indicate the overall cost and carbon emissions with the Z pole and WW threshold runs of FCC-ee and with the tunnel(s) and infrastructure of the 10 TeV pCM collider following the linear Higgs factory.}
    \begin{tabular}{c|c|c|c}
    \hline 
     Collider &  FCC-ee$_{240}$ &  CLIC$_{380}$ & ILC@CERN$_{250}$ \\ \hline
         Operation time (years) & 5 & 47  & 45 \\ \hline
         Electricity consumption (TWh) & 6.7  & 28.2 & 31.0 \\ \hline
         Cost (Billion euros) & 18.7  & 27.1  & 30.1 \\ \hline
         Carbon budget (MtCO$_2$e) & 0.66  & 0.64 & 0.78 \\ \hline\hline         
         Overall cost (Billion euros) & 23.7  & 35.3 & 38.3\\ \hline
         Overall carbon budget (MtCO$_2$e) & 0.79  & 1.05 & 1.19 \\ \hline         
    \end{tabular}
\end{table}

At FCC-ee, for example, a short run to scan the $\rm t \bar{t}$ threshold ($\sqrt{s} = 340$--350\,GeV) allows the measurement of the top-quark mass, a fundamental parameter of the Standard Model, with a statistical precision of $\cal O$(10\,MeV). Because the prediction of electroweak precision observables (EWPO) is, in various ways, sensitive to $m_{\rm top}$, the discovery power of this EWPO exploration is limited by the uncertainty on $m_{\rm top}$. For example, matching the precision of the SM predictions from the EWPO measurements to the 180\,keV (resp.\ 4\,keV) statistical uncertainty on the W mass (resp.\ Z width) requires a 20\,MeV (resp.\ 15\,MeV) knowledge of $m_{\rm top}$.  
Moreover, the ZH cross section dependence on $\sqrt{s}$ provides sensitivity to the Higgs boson self-coupling when data at $\sqrt{s} = 365$\,GeV are available, as alluded to in Section~\ref{sec:self}. More importantly, the per-cent measurement of the top EW couplings at $365$\,GeV {\it(i)}~matches the EWPO ppm precision at the Z pole and the WW threshold; and {\it (ii)}~keep the theoretical uncertainties on the top Yukawa coupling determination at the FCC-hh at the per-cent level, a pre-requisite for the model-independent per-cent determination of the Higgs boson self-coupling at FCC-hh.

\subsection{Luminosity upgrade}

\begin{figure}[htbp]
\centering
\includegraphics[width=0.62\textwidth]{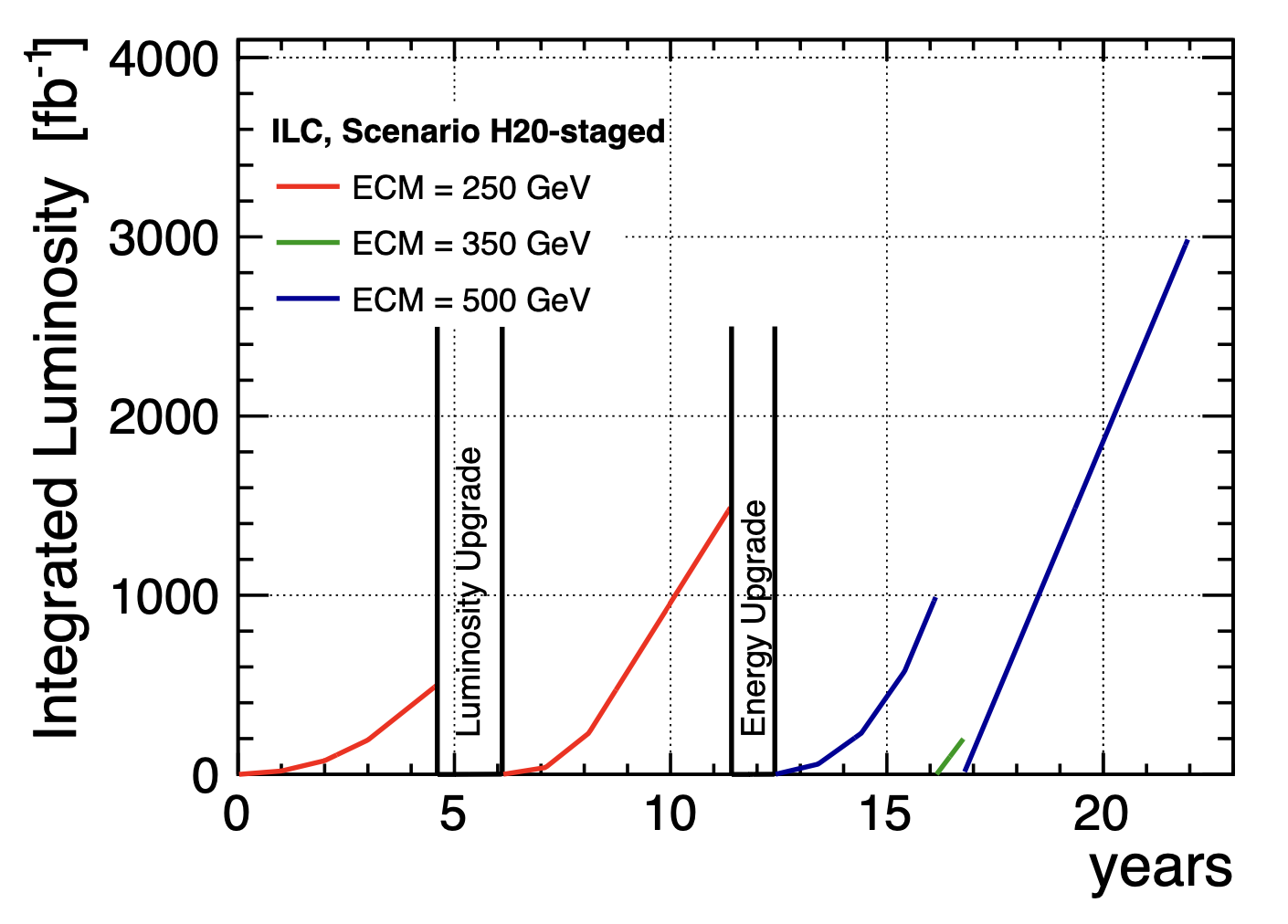}
\caption{\small The plan for operations of ILC through its various stages from 250 to 500\,GeV, that is used in Ref.~\cite{ILCInternationalDevelopmentTeam:2022izu} for projections of the physics results used in the fit of Ref.~\cite{deBlas:2022ofj}. In this plan, the time for collisions is assumed to be $1.6\,10^7$ seconds per year.  
\label{fig:ILCRunPLan}
}
\end{figure}

The current run plan of ILC (Fig.~\ref{fig:ILCRunPLan}) is to start at 250\,GeV with an instantaneous luminosity of $1.35\times 10^{34}$ cm$^{-2}$s$^{-1}$, and double it after five years (plus one year shutdown to install the upgrade) with twice more bunches per RF pulse. The possibility that ILC could start right away with $2.7\times 10^{34}$ cm$^{-2}$s$^{-1}$ is now being contemplated.

With twice larger instantaneous luminosity, the integrated luminosity would also be doubled after 4.5 years: 0.9\,ab$^{-1}$ instead of 0.45\,ab$^{-1}$. At CERN, as already mentioned, years are slightly shorter than what is assumed in Ref.~\cite{ILCInternationalDevelopmentTeam:2022izu}: $1.2\times 10^7$ seconds instead of $1.6\times 10^7$ seconds, making it 6 years to get to 0.9\,ab$^{-1}$. With 0.324\,ab$^{-1}$ per year after these 6 years, it would take another 3.5 years to reach 2\,ab$^{-1}$, thus reducing the initial 15 years to about 10 years for the default ILC@CERN run at 250\,GeV, keeping all the rest unmodified. All operation times presented for ILC in the previous two sections would then be shortened by 5 years (25 years instead of 30 years at 250\,GeV to reach the coupling precisions that FCC-ee$_{240}$ achieves in 3 years, and 41 (40) years instead of 46 (45) to reach the precisions that FCC-ee$_{240+365}$ (FCC-ee$_{240}$) achieves in 8 (5) years. The cost and the carbon emissions of the 250\,GeV stage would shrink by 17\% (i.e., a reduction of less than 3\% of the total cost and less than 5\% of the total carbon emissions). The main conclusions drawn in this note would remain unchanged.

\subsection{Pushing the limits}
\label{sec:power}

More aggressive ways to double the instantaneous luminosities have been suggested for FCC-ee$_{240}$~\cite{frankAnnecy} and ILC$_{250}$~\cite{jennyDesy}.  For FCC-ee, this can be achieved either by increasing the number of bunches by (thus increasing the synchrotron radiation (SR) power in the same proportions) and/or by decreasing the vertical $\beta^\ast$ (along with the vertical emittance). If the number of bunches were increased by 50\%, the total energy consumption would then increase by about 30\% and would become similar to that of the t\=t stage, i.e., 1.77\,TWh/year. These new settings would come with their own challenges, but no hardware would need to be modified with respect to the baseline layout. For ILC, the doubling of instantaneous luminosity at 250\,GeV would require, e.g., to build the 500\,GeV collider right from the beginning (to have enough RF power to increase the repetition rate from 5\,Hz to 10\,Hz), to upgrade already challenging damping rings, to develop a kicker able to deal with 3\,ns bunch spacing and critical collective effects, and to redesign the cryo-modules and their distribution~\cite{torpriv}. A costing of these modifications is not yet available. The total energy consumption at 250\,GeV would almost double and equal that of the 500\,GeV configuration, i.e., 1.2\,TWh/year.   

{With these modifications, it would take a little over 3 years with 0.648\,\,ab$^{-1}$/year to accumulate 2\,ab$^{-1}$ at 250\,GeV, and reach a precision of 0.37\% on the Higgs coupling to the Z boson (Table~\ref{tab:SMEFT250}), extending to almost 20 years to reach the same precision as FCC-ee in three years at 240\,GeV. As the luminosity for the second stage  would not change, it would still take a little over 12 years for ILC@CERN to accumulate the desired integrated luminosities at 350 and 500\,GeV (0.1 and 4\,ab$^{-1}$, respectively, in agreement with the projections in Ref.~\cite{jennyDesy}), and reach a combined precision of 0.26\% on the Higgs coupling to the Z boson (Table~\ref{tab:SMEFT500}). The precision expected after 8 years at FCC-ee, with twice larger luminosity at 240\,GeV, would be 0.13\%, i.e., twice better than after $12+3=15$ years of ILC@CERN. For ILC@CERN to reach the same precision as FCC-ee in 8 years, four times more integrated luminosity would be needed, achievable in 12 years at 250\,GeV and 40 years at 500\,GeV, for a total of 52 years. The corresponding electricity consumptions, carbon emissions and cost (without including the cost and carbon budget of the additional hardware needed for ILC) are summarised in Table~\ref{tab:stretch}.

\begin{table}[h]
\renewcommand{\arraystretch}{1.25}
\centering
\caption{\label{tab:stretch} \small Operation time, electricity consumption, construction and operation cost and carbon budget, for FCC-ee$_{240+365}$ and ILC@CERN$_{250+500}$ to achieve the same Higgs coupling precision as the planned FCC-ee$_{240+365}$ runs, when doubling the instantaneous luminosity in the first stage (see text). The last two rows indicate, for illustration,  the overall cost and carbon emissions with the Z pole and WW threshold runs of FCC-ee and with the tunnel(s) and infrastructure of the 10 TeV pCM collider following the linear Higgs factory.}
    \begin{tabular}{c|c|c}
    \hline 
     Collider &  FCC-ee$_{240+365}$ & ILC@CERN$_{250+500}$ \\ \hline
         Operation time (years) & 8 & 52 \\ \hline
         Electricity consumption (TWh) & 14.1 & 61.0 \\ \hline
         Cost (Billion euros) & 23.5 & 52.8 \\ \hline
         Carbon budget (MtCO$_2$e) & 0.81 & 1.46 \\ \hline\hline         
         Overall cost (Billion euros) & 28.5 & 61.0\\ \hline
         Overall carbon budget (MtCO$_2$e) & 0.94 & 1.87 \\ \hline         
    \end{tabular}
\end{table}

Given the existing challenges to deliver the design luminosity in all projects, however, it would seem more prudent -- using past experience as a guide -- to assume that the luminosities might be smaller than their design values.

\subsection{Caveat and uncertainties}
\label{sec:caveat}

As already mentioned, the Higgs coupling precisions are assumed to improve as fast as the square root of the integrated luminosity of the considered $\rm e^+e^-$ colliders. The actual improvement is slightly slower as the $\rm e^+e^-$ collider inputs are combined with HL-LHC in the fits of Ref.~\cite{deBlas:2022ofj}. The effect of this combination is minimized here by limiting the analysis to the five couplings dominated by $\rm e^+e^-$ colliders: Z, W, b, c and $\tau$, but is not entirely suppressed. With the default operation models assumed in Ref.~\cite{deBlas:2022ofj}, the bootstrapping FCC-ee precision are better than the ILC precision, and the ILC precision are better than the CLIC precision. This caveat is thus favourable to CLIC and, to a lesser extent, to ILC in the conclusions of the present work. 

This assumption of a precision improvement as fast as the square root of the integrated luminosity would also be invalidated if experimental systematic uncertainties in excess of the per-mil level were to saturate statistical uncertainties. Experimental systematic uncertainties, however, are very often of statistical nature. At FCC-ee, it is planned to collect $10^9$ Z decays at $\sqrt{s} = 91.2$\,GeV for a few hours every month, which would allow these statistics-limited experimental systematic uncertainties to be tamed well below a per mil. At linear colliders, with a three-order of magnitude smaller luminosity at the Z pole, experimental systematic uncertainties might not be reducible below a per cent. Events at $\sqrt{s} = 250$\,GeV with a radiative return to the Z or with a pair of Z are not more numerous, and such events at $\sqrt{s} = 500$\,GeV are too forward-boosted to be entirely useful for detector calibration and alignment. This assumption is therefore again very much favourable to linear colliders in this note's conclusions.    

Finally, the coupling precisions listed in Ref.~\cite{deBlas:2022ofj} and used in the present work are rounded to the second digit, and are thus affected by a systematic bias of up to $\pm 5\%$. When propagated to the operation times (which vary like the square of the coupling precisions) inferred in Section~\ref{sec:duration} for linear colliders, this systematic bias amounts to $\pm 10\%$, i.e., $\pm 5$ years for the 50 years operation of CLIC and ILC@CERN, which translates to $\pm 5$\,TWh (out of 50) for their electricity consumption, $\pm 0.1$\,Mt CO$_2$e (out of 1.5) for their carbon budget, and $\pm 4$ billion euros (out of 60) for their cost.

\subsection{Half a century}
\label{sec:50}

Running low-luminosity colliders for half a century, possibly with only one interaction point, may not be a particularly attractive option, for several reasons. A few of these reasons are listed here. 
\begin{itemize}
\item Scientifically, it may well be that Higgs-coupled new physics in the 10\,TeV region gives a $3\sigma$ deviation on the Higgs coupling to the Z (the best measured coupling of all) after the eight years of the FCC-ee running. A confirmation of such an evidence would require another eight years at FCC-ee, which is a manageable time quantum. Instead, it would be unthinkable to run for a whole century with ILC@CERN and CLIC for such a confirmation.
\item Pragmatically, if the luminosity happens to be smaller than expected by a factor 2 and a per-mil precision is aimed at on the Higgs coupling to the Z boson, it is in principle possible at FCC-ee to schedule a twice longer Higgs run without grossly distorting the overall timeline for particle physics projects, but it would be impossible with linear colliders. 
\item Technologically, it is very difficult to imagine that the hardware equipment, detector and accelerator alike, survives half a century of continuous operation. Some of the LEP detectors were already falling apart after 11 years of running, and it would have been hard to continue operating the accelerating RF cavities for much longer.  
\item  Humanly, it would also not particularly motivating for high-energy physicists, as such a long time would exceed their professional lifetime, for a rather limited set of measurements.
\item Independently of the delay caused by these 50 years of operation for the putative start of FCC-hh, building a linear collider at CERN would annihilate the perspective of a large hadron circular collider at the same site: the Host States would not tolerate the construction -- and all the burden thereof -- of two major colliders in their territory.  
\end{itemize}

A frequently heard  counter-argument is that the ILC is ``shovel-ready'' and therefore could start earlier than FCC-ee; and that the six years needed for the Z and WW runs at FCC-ee would further delay the Higgs run. The first statement may have been true a decade ago in Japan, had a timely decision been taken but, at CERN, no new collider can start before the mid 2040's, a few years after the end of HL-LHC. Moreover, following a recommendation from the FCC feasibility study mid-term review committees to {\it consolidate (and ideally simplify) the design of the RF system to allow efficient energy-staging, as well as to reduce complexity, risk, and cost; and to study options to avoid the 1-cell/2-cell RF cavity reconfiguration between Z and ZH/WW running, in order to simplify the RF system implementation and to improve flexibility in the physics programme}, a versatile RF scheme has been designed that enables a quasi-total flexibility to choose the running sequence. The 240\,GeV run could then proceed early, before going back to the Z pole and the WW threshold: much more will be learned about the Higgs boson, and much quicker, with FCC-ee.

\subsection{Comparison to earlier work}
\label{sec:comparison}

The conclusion of Section~\ref{sec:overview} is qualitatively no different from what was reported in the very first analysis of Ref.~\cite{Janot:2022jtn}, but is arguably a lot more robust. With respect to this earlier analysis, the added value of the work reported here includes {\it(i)} the life cycle assessments of the collider construction carbon budget and cost (previously unavailable); {\it (ii)} a projection for the carbon intensity of the electricity production at the time of the collider operation (instead of using today's figures); {\it (iii)} a better comparison of the Higgs factory physics outcomes (directly with the Higgs coupling precision instead of the number of Higgs boson produced); and {\it(iv)} a full account of the 
beam longitudinal polarisation gains in the SMEFT coupling fit of Ref.~\cite{deBlas:2022ofj}.

Instead, Refs.~\cite{Breidenbach:2023nxd,Bloom:2024ixe} both reach the opposite conclusion for the environmental impact, and disagree on the energy consumption. A critical review of these two notes revealed the following inaccuracies in their reasoning and bibliography. 
\begin{enumerate}
    \item Both papers use today's figures for the civil engineering carbon footprint but projected estimates for the carbon intensity of electricity production. 
    \vskip 0.05cm
    \item In both papers, the carbon budget of the second stage of CLIC (civil engineering, infrastructures, and operation) and ILC (civil engineering and infrastructures) is not included. The grand-vision of a 10 TeV pCM collider, as articulated by Snowmass/P5 and ESU2020, is ignored. \vskip 0.05cm
    \item In Ref.~\cite{Bloom:2024ixe}, the carbon emissions of the collider operation is assumed to be zero as of 2040 (probably because of a confusion with the net-zero emissions imposed by EU, in which actual emissions are neutralised by purchasing emission offsets~\cite{robipriv}). This assumption is overly optimistic and is also not permitted according to the EU guidelines~\cite{jopriv}. Even if electricity were to be produced entirely by renewable energy sources, the carbon emissions of the development and the operation of these sources would still be of the order of 20\,kg CO$_2$e per MWh. As a matter of fact, this carbon intensity value was chosen both in Ref.~\cite{Breidenbach:2023nxd} and for the present work. \vskip 0.05cm
    \item In none of the two papers is the concept of operation time considered relevant. It does not matter in Ref.~\cite{Bloom:2024ixe} as the operation gross emissions are unrealistically assumed to vanish after 2040.  In Ref.~\cite{Breidenbach:2023nxd}, the absolute operation emissions are limited to the default run plans (20 years for ILC and 8 years for CLIC). \vskip 0.05cm
    \item Finally, in Ref.~\cite{Breidenbach:2023nxd}, an attempt to include the coupling precisions as a multiplicative (reduction) factor in the carbon emissions was made, with goal of obtaining a ``carbon footprint per unit of physics output''. This attempt is flawed and leads to distorted conclusions, as explained in details in Ref.~\cite{janot_2024_gbp32-1b445}. 
\end{enumerate}

These inaccuracies bias the results by large cumulative factors and in only one direction, namely that of increasing the FCC-ee Higgs factory carbon budget with respect to that of ILC and CLIC Higgs factories (items 1 and 5), and decreasing the carbon budgets of ILC and CLIC with respect to that of FCC-ee (items 2, 3, 4 and 5).

\section{Concluding remarks}
\label{sec:conclusion}

The choice of the next collider at CERN after the LHC will be guided, in particular, by two important features of the human beings: their finite lifetime and their ability to dream. This new collider will therefore have to push the limits of the unknown as far and as fast as possible, in the broadest possible way, and with a real chance of discovery, in an entirely new context where neither the mass scale of new physics nor the intensity of its couplings to the Standard Model are known. 

Pragmatically, we confirmed once again in this paper that what truly matters is integrated luminosity and integrated energy consumption (as opposed to instantaneous power), especially for a ``factory'' machine that does not run at its maximum possible energy. With the highest luminosities at the electroweak scale and parton-parton collision energies an order of magnitude above those of the LHC, the fully synergistic FCC integrated programme, with a versatile ${\rm e^+e^-}$ collider in a first 15 years stage, followed by 25 years of 100\,TeV proton-proton collisions, optimises the overall investment and the breadth of its science value, by addressing the following diverse and ambitious goals: 
\begin{itemize}
    \item Map the properties of the Higgs and electroweak gauge bosons, pinning down their interactions with accuracies order(s) of magnitude better than today, and acquiring sensitivity to, e.g.\ the processes that led to the formation of today's Higgs vacuum field right after the Big Bang.
    \item  Sharpen the knowledge of already identified particle physics phenomena with a comprehensive campaign of precision electroweak, QCD, flavour, tau, Higgs, and top measurements, sensitive to tiny deviations from the predicted Standard Model behaviour and probing energy scales far beyond the direct kinematic reach. 
    \item Improve by orders of magnitude the sensitivity to rare and elusive phenomena at low energies, including the possible discovery of light particles with very small couplings. In particular, the search for dark matter should seek to reveal, or conclusively exclude, dark sector  candidates belonging to broad classes of models.
    \item Improve, by at least an order of magnitude, the direct discovery reach for new particles at the energy frontier. A proton-proton collider has a unique virtue in this respect, as it allows the sampling of a wide variety of initial states, from gluons to photons or EW gauge bosons, and quarks of all flavours, and from neutral to charged or even coloured initial states, any of which could lead to the production of a new particle. 
\end{itemize}

Most of the above aspects are simply inaccessible to linear $\rm e^+e^-$ colliders. In this note, only the few measurements of the Higgs boson properties also accessible to linear colliders, have been used as a common platform for comparison. Much more will be learned about the Higgs boson with FCC (such as couplings to $\gamma\gamma$, $\rm Z\gamma$, $\mu^+\mu^-$, much better measured at FCC-hh; and to $\rm e^+e^-$ and $\rm s\bar s$, uniquely accessible at FCC-ee), and much quicker. A precise measurement of the Higgs self-coupling at the few per-cent precision level can realistically only be provided by the combination of FCC-ee and FCC-hh, which is beyond the reach of lepton colliders with centre-of-mass energies up to 3\,TeV. 

The range extending from 0.4 to 2.5\,TeV, in which linear ${\rm e^+e^-}$ colliders could have direct sensitivity to new physics is already broadly covered by LHC.\footnote{Unnatural regions of parameter space with specific cancellations or tunings may accidentally escape this broad coverage.} As a matter of fact, the original motivation (from B. Richter in 1976) to reach high centre-of-mass energies with linear $\rm e^+e^-$ colliders, was developed at a time when the top quark and Higgs boson masses, and the fact that there are only three families of standard-model fermions, were not known. Since then, all new heavy particles (b quark, Z and W bosons, top quark, Higgs boson) have been discovered exclusively through hadron collisions.
In addition, as muon colliders need circular tunnels, they would not be a ``natural'' evolution of any linear ${\rm e^+e^-}$  collider facility. In contrast, the CERN infrastructure formed by two rings of 27\,km (LEP/LHC) and 91\,km (FCC-ee/FCC-hh) would offer the synergistic opportunity to study a very-high energy muon collider in the LEP/LHC tunnel, with a fast muon acceleration system in the FCC tunnel. 

Time has come to get together and develop innovative and ambitious technical solutions (accelerator and detectors alike) and to tackle all FCC challenges in front of us. In particular, although FCC already promises the broadest scientific outcome for a given operation time, electricity consumption, carbon footprint, and cost, we are still working actively to improve on these aspects, for both FCC-ee and FCC-hh. 

\section*{Disclaimer}
The results presented in this note are complementary to the ongoing work of the European Laboratory Directors Group (LDG), the CERN Future Colliders Comparative Evaluation Working Group, or the CERN Sustainable Accelerators Panel. These groups are more into evaluating the absolute electricity consumption and carbon emissions for each collider, under the implicit assumption that all colliders are more or less scientifically interchangeable. The results presented in this note, instead, evaluate the same quantities {\it for a given physics outcome} and use these evaluations as an unbiased sustainability metric. The cost and carbon footprint for the construction of the various options studied in this note, while more complete than those in Refs.~\cite{Bloom:2024ixe,Breidenbach:2023nxd}, still miss a few contributions. The current estimates are accurate and complete enough for the comparison carried out here, but this note will be updated as soon as official figures are made public. 

\section*{Acknowledgements}

Special thanks are due to
Michael Benedikt,
Jorge de Blas,
David d'Enterria,
Fabiola Gianotti,
Jiayin Gu,
Marumi Kado,
Roberto Losito,
Carlos Louren\c{c}o,
Michelangelo Mangano, 
Srini Rajagopalan,
Tor Raubeheimer,
Michele Selvaggi,
Florian Sonnemann,
and Frank Zimmermann
for their useful comments and suggestions before and during the preparation of this note. This work has been partially funded from the European Union's Horizon 2020 research and innovation programme under grant agreement No. 951754. and is supported by the Deutsche Forschungsgemeinschaft under Germany’s Excellence Strategy EXC 2121 ``Quantum Universe'' -- 390833306, as well as by the grant 491245950. This project also has received funding from the European Union’s Horizon Europe research and innovation programme under the Marie Sk\l{}odowska-Curie Staff Exchange grant agreement No 101086085 - ASYMMETRY. 

\bibliography{References}
\end{document}